\documentclass[twocolumn,showpacs,superscriptaddress,altaffilletter]{revtex4}

\usepackage{graphicx}
\usepackage{dcolumn}
\usepackage{amsmath}

\def\kevc1{\ifmmode\mathrm{\ keV/{\mit c}}
          \else$\mathrm{\ keV/{\mit c}}$\fi}
\def\Mevc1{\ifmmode\mathrm{\ MeV/{\mit c}}
          \else$\mathrm{\ MeV/{\mit c}}$\fi}
\def\gevc1{\ifmmode\mathrm{\ GeV/{\mit c}}
          \else$\mathrm{\ GeV/{\mit c}}$\fi}
\def\kevc2{\ifmmode\mathrm{\ keV/{\mit c}^2}
          \else$\mathrm{\ keV/{\mit c}^2}$\fi}
\def\Mevc2{\ifmmode\mathrm{\ MeV/{\mit c}^2}
          \else$\mathrm{\ MeV/{\mit c}^2}$\fi}
\def\Gevc2{\ifmmode\mathrm{\ GeV/{\mit c}^2}
          \else$\mathrm{\ GeV/{\mit c}^2}$\fi}
\def\Gev2c2{\ifmmode\mathrm{\ GeV^2/{\mit c}^2}
          \else$\mathrm{\ GeV^2/{\mit c}^2}$\fi}

\def\ubar{\ifmmode\mathrm{\overline {u}}
          \else$\mathrm{\overline{u}}$\fi}
\def\dbar{\ifmmode\mathrm{\overline {d}}
          \else$\mathrm{\overline{d}}$\fi}
\def\sbar{\ifmmode\mathrm{\overline {s}}
          \else$\mathrm{\overline{s}}$\fi}
\def\cbar{\ifmmode\mathrm{\overline {c}}
          \else$\mathrm{\overline{c}}$\fi}
\def\bbar{\ifmmode\mathrm{\overline {b}}
          \else$\mathrm{\overline{b}}$\fi}
\def\tbar{\ifmmode\mathrm{\overline {t}}
          \else$\mathrm{\overline{t}}$\fi}
\def\qbar{\ifmmode\mathrm{\overline {q}}
          \else$\mathrm{\overline{q}}$\fi}

\def\uq{\ifmmode\mathrm{u}
          \else$\mathrm{u}$\fi}
\def\dq{\ifmmode\mathrm{d}
          \else$\mathrm{d}$\fi}
\def\sq{\ifmmode\mathrm{s}
          \else$\mathrm{s}$\fi}
\def\cq{\ifmmode\mathrm{c}
          \else$\mathrm{c}$\fi}
\def\bq{\ifmmode\mathrm{b}
          \else$\mathrm{b}$\fi}
\def\tq{\ifmmode\mathrm{t}
          \else$\mathrm{t}$\fi}
\def\qq{\ifmmode\mathrm{q}
          \else$\mathrm{q}$\fi}

 \def\Pgg{\ifmmode\mathrm{\gamma}
          \else$\mathrm{\gamma}$\fi}
 \def\PW{\ifmmode\mathrm{W}
         \else$\mathrm{W }$\fi}
 \def\PWp{\ifmmode\mathrm{W^+}
          \else$\mathrm{W^+}$\fi}
 \def\PWpm{\ifmmode\mathrm{W^{\pm}}
          \else$\mathrm{W^{\pm}}$\fi}
 \def\PWm{\ifmmode\mathrm{W^-}
          \else$\mathrm{W^-}$\fi}
 \def\PZz{\ifmmode\mathrm{Z^0}
          \else$\mathrm{Z^0}$\fi}
 \def\PHz{\ifmmode\mathrm{H^0}
          \else$\mathrm{H^0}$\fi}
 \def\PHpm{\ifmmode\mathrm{H^{\pm}}
           \else$\mathrm{H^{\pm}}$\fi}
 \def\PWR{\ifmmode\mathrm{W_R}
          \else$\mathrm{W_R}$\fi}
 \def\PWpr{\ifmmode\mathrm{W^{\prime}}
           \else$\mathrm{W^{\prime}}$\fi}
 \def\PZLR{\ifmmode\mathrm{Z_{LR}}
           \else$\mathrm{Z_{LR}}$\fi}
 \def\PZgc{\ifmmode\mathrm{Z_{\chi}}
           \else$\mathrm{Z_{\chi}}$\fi}
 \def\PZgy{\ifmmode\mathrm{Z_{\psi}}
           \else$\mathrm{Z_{\psi}}$\fi}
 \def\PZge{\ifmmode\mathrm{Z_{\eta}}
           \else$\mathrm{Z_{\eta}}$\fi}
 \def\PZi{\ifmmode\mathrm{Z_1}
          \else$\mathrm{Z_1}$\fi}
 \def\PAz{\ifmmode\mathrm{A^0}
          \else$\mathrm{A^0}$\fi}
 \def\Pgne{\ifmmode\mathrm{\nu_{e}}
           \else$\mathrm{\nu_{e}}$\fi}
 \def\Pagne{\ifmmode\mathrm{\overline{\nu_{e}}}
            \else$\mathrm{\overline{\nu_{e}}}$\fi}
 \def\Pgngm{\ifmmode\mathrm{\nu_{\mu}}
            \else$\mathrm{\nu_{\mu}}$\fi}
 \def\Pagngm{\ifmmode\mathrm{\overline{\nu}_{\mu}}
             \else$\mathrm{\overline{\nu}_{\mu}}$\fi}
 \def\Pgngt{\ifmmode\mathrm{\nu_{\tau}}
            \else$\mathrm{\nu_{\tau}}$\fi}
 \def\Pagngt{\ifmmode\mathrm{\overline{\nu}_{\tau}}
             \else$\mathrm{\overline{\nu}_{\tau}}$\fi}
 \def\Pe{\ifmmode\mathrm{e}
         \else$\mathrm{e}$\fi}
 \def\Pep{\ifmmode\mathrm{e^+}
          \else$\mathrm{e^+}$\fi}
 \def\Pem{\ifmmode\mathrm{e^-}
          \else$\mathrm{e^-}$\fi}
 \def\Pgm{\ifmmode\mathrm{\mu}
          \else$\mathrm{\mu}$\fi}
 \def\Pgmm{\ifmmode\mathrm{\mu^-}
           \else$\mathrm{\mu^-}$\fi}
 \def\Pgmp{\ifmmode\mathrm{\mu^+}
           \else$\mathrm{\mu^+}$\fi}
 \def\Pgt{\ifmmode\mathrm{\tau}
          \else$\mathrm{\tau}$\fi}
 \def\PLpm{\ifmmode\mathrm{L^{\pm}}
           \else$\mathrm{L^{\pm}}$\fi}
 \def\PLz{\ifmmode\mathrm{L^0}
          \else$\mathrm{L^0}$\fi}
 \def\PEz{\ifmmode\mathrm{E^0}
          \else$\mathrm{E^0}$\fi}
 \def\Pgp{\ifmmode\mathrm{\pi}
          \else$\mathrm{\pi }$\fi}
 \def\Pgpm{\ifmmode\mathrm{\pi^-}
           \else$\mathrm{\pi^-}$\fi}
 \def\Pgpp{\ifmmode\mathrm{\pi^+}
           \else$\mathrm{\pi^+}$\fi}
 \def\Pgppm{\ifmmode\mathrm{\pi^{\pm }}
            \else$\mathrm{\pi^{\pm }}$\fi}
 \def\Pgpz{\ifmmode\mathrm{\pi^0}
           \else$\mathrm{\pi^0 }$\fi}
 \def\Pgh{\ifmmode\mathrm{\eta}
          \else$\mathrm{\eta }$\fi}
 \def\Pgr{\ifmmode\mathrm{\rho(770)}
          \else$\mathrm{\rho(770)}$\fi}
 \def\Pgo{\ifmmode\mathrm{\omega(783)}
          \else$\mathrm{\omega(783)}$\fi}
 \def\Pghpr{\ifmmode\mathrm{\eta^{\prime}(958)}
            \else$\mathrm{\eta^{\prime}(958)}$\fi}
 \def\Pfz{\ifmmode\mathrm{f_0(980)}
          \else$\mathrm{f_0(980)}$\fi}
 \def\Paz{\ifmmode\mathrm{a_0(980)}
          \else$\mathrm{a_0(980)}$\fi}
 \def\Pgf{\ifmmode\mathrm{\phi(1020)}
          \else$\mathrm{\phi(1020)}$\fi}
 \def\Phia{\ifmmode\mathrm{h_1(1170)}
           \else$\mathrm{h_1(1170)}$\fi}
 \def\Pbi{\ifmmode\mathrm{b_1(1235)}
          \else$\mathrm{b_1(1235)}$\fi}
 \def\Pai{\ifmmode\mathrm{a_1(1260)}
          \else$\mathrm{a_1(1260)}$\fi}
 \def\Pfii{\ifmmode\mathrm{f_2(1270)}
           \else$\mathrm{f_2(1270)}$\fi}
 \def\Pfi{\ifmmode\mathrm{f_1(1285)}
          \else$\mathrm{f_1(1285)}$\fi}
 \def\Pgha{\ifmmode\mathrm{\eta(1295)}
           \else$\mathrm{\eta(1295)}$\fi}
 \def\Pgpa{\ifmmode\mathrm{\pi(1300)}
           \else$\mathrm{\pi(1300)}$\fi}
 \def\Paii{\ifmmode\mathrm{a_2(1320)}
           \else$\mathrm{a_2(1320)}$\fi}
 \def\Pgoa{\ifmmode\mathrm{\omega(1390)}
           \else$\mathrm{\omega(1390)}$\fi}
 \def\Pfza{\ifmmode\mathrm{f_0(1400)}
           \else$\mathrm{f_0(1400)}$\fi}
 \def\Pfia{\ifmmode\mathrm{f_1 (1390)}
           \else$\mathrm{f_1 (1390)}$\fi}
 \def\Pghb{\ifmmode\mathrm{\eta(1440)}
           \else$\mathrm{\eta(1440)}$\fi}
 \def\Pgra{\ifmmode\mathrm{\rho(1450)}
           \else$\mathrm{\rho(1450)}$\fi}
 \def\Pfib{\ifmmode\mathrm{f_1(1510)}
           \else$\mathrm{f_1(1510)}$\fi}
 \def\Pfiipr{\ifmmode\mathrm{f^{\prime}_2(1525)}
             \else$\mathrm{f^{\prime}_2(1525)}$\fi}
 \def\Pfzb{\ifmmode\mathrm{f_0(1590)}
           \else$\mathrm{f_0(1590)}$\fi}
 \def\Pgob{\ifmmode\mathrm{\omega(1600)}
           \else$\mathrm{\omega(1600)}$\fi}
 \def\Pgoiii{\ifmmode\mathrm{\omega_3(1670)}
             \else$\mathrm{\omega_3(1670)}$\fi}
 \def\Pgpii{\ifmmode\mathrm{\pi_2(1670)}
            \else$\mathrm{\pi_2(1670)}$\fi}
 \def\Pgfa{\ifmmode\mathrm{\phi(1680)}
           \else$\mathrm{\phi(1680)}$\fi}
 \def\Pgriii{\ifmmode\mathrm{\rho_3(1690)}
             \else$\mathrm{\rho_3(1690)}$\fi}
 \def\Pgrb{\ifmmode\mathrm{\rho(1700)}
           \else$\mathrm{\rho(1700)}$\fi}
 \def\Pfiia{\ifmmode\mathrm{f_2(1720)}
            \else$\mathrm{f_2(1720)}$\fi}
 \def\Pgfiii{\ifmmode\mathrm{\phi_3(1850)}
             \else$\mathrm{\phi_3(1850)}$\fi}
 \def\Pfiib{\ifmmode\mathrm{f_2(2010)}
            \else$\mathrm{f_2(2010)}$\fi}
 \def\Pfiv{\ifmmode\mathrm{f_4(2050)}
           \else$\mathrm{f_4(2050)}$\fi}
 \def\Pfiic{\ifmmode\mathrm{f_2(2300)}
            \else$\mathrm{f_2(2300)}$\fi}
 \def\Pfiid{\ifmmode\mathrm{f_2(2340)}
            \else$\mathrm{f_2(2340)}$\fi}
 \def\PK{\ifmmode\mathrm{K}
         \else$\mathrm{K}$\fi}
 \def\PKpm{\ifmmode\mathrm{K^{\pm}}
           \else$\mathrm{K^{\pm}}$\fi}
 \def\PKp{\ifmmode\mathrm{K^+}
          \else$\mathrm{K^+}$\fi}
 \def\PKm{\ifmmode\mathrm{K^-}
          \else$\mathrm{K^-}$\fi}
 \def\PKz{\ifmmode\mathrm{K^0}
          \else$\mathrm{K^0}$\fi}
 \def\PaKz{\ifmmode\mathrm{\overline{K^0}}
           \else$\mathrm{\overline{K^0}}$\fi}
 \def\PKgmiii{\ifmmode\mathrm{K_{\mu 3}}
              \else$\mathrm{K_{\mu 3}}$\fi}
 \def\PKeiii{\ifmmode\mathrm{K_{\rm e3}}
             \else$\mathrm{K_{\rm e3}}$\fi}
 \def\PKzS{\ifmmode\mathrm{K^0_{\rm S}}
           \else$\mathrm{K^0_{\rm S}}$\fi}
 \def\PKzL{\ifmmode\mathrm{K^0_{\rm L}}
           \else$\mathrm{K^0_{\rm L}}$\fi}
 \def\PKzgmiii{\ifmmode\mathrm{K^0_{\mu 3}}
               \else$\mathrm{K^0_{\mu 3}}$\fi}
 \def\PKzeiii{\ifmmode\mathrm{K^0_{{\rm e}3}}
              \else$\mathrm{K^0_{{\rm e}3}}$\fi}
 \def\PKst{\ifmmode\mathrm{K^{\ast}(892)}
           \else$\mathrm{K^{\ast}(892)}$\fi}
 \def\PKi{\ifmmode\mathrm{K_1(1270)}
          \else$\mathrm{K_1(1270)}$\fi}
 \def\PKsta{\ifmmode\mathrm{K^{\ast}(1370)}
            \else$\mathrm{K^{\ast}(1370)}$\fi}
 \def\PKia{\ifmmode\mathrm{K_1(1400)}
           \else$\mathrm{K_1(1400)}$\fi}
 \def\PKstz{\ifmmode\mathrm{K^{\ast}_0(1430)}
            \else$\mathrm{K^{\ast}_0(1430)}$\fi}
 \def\PKstii{\ifmmode\mathrm{K^{\ast}_2(1430)}
             \else$\mathrm{K^{\ast}_2(1430)}$\fi}
 \def\PKstb{\ifmmode\mathrm{K^{\ast}(1680)}
            \else$\mathrm{K^{\ast}(1680)}$\fi}
 \def\PKii{\ifmmode\mathrm{K_2(1770)}
           \else$\mathrm{K_2(1770)}$\fi}
 \def\PKstiii{\ifmmode\mathrm{K^{\ast}_3(1780)}
              \else$\mathrm{K^{\ast}_3(1780)}$\fi}
 \def\PKstiv{\ifmmode\mathrm{K^{\ast}_4(2045)}
             \else$\mathrm{K^{\ast}_4(2045)}$\fi}
 \def\PD{\ifmmode\mathrm{D}
           \else$\mathrm{D}$\fi}
 \def\PaD{\ifmmode\mathrm{\overline{ D}}
          \else${\mathrm{\overline D}}$\fi}
 \def\PDpm{\ifmmode\mathrm{D^{\pm}}
           \else$\mathrm{D^{\pm}}$\fi}
 \def\PDm{\ifmmode\mathrm{D^-}
          \else$\mathrm{D^-}$\fi}
 \def\PDp{\ifmmode\mathrm{D^+}
          \else$\mathrm{D^+}$\fi}
 \def\PDz{\ifmmode\mathrm{D^0}
          \else$\mathrm{D^0}$\fi}
 \def\PaDz{\ifmmode\mathrm{\overline{D^0}}
           \else$\mathrm{\overline{D^0}}$\fi}
 \def\PDstpm{\ifmmode{\mathrm{D}^{\ast}(2010)^{\pm}}
             \else$\mathrm{D}^{\ast}(2010)^{\pm}$\fi}
 \def\PDstp{\ifmmode{\mathrm{D}^{\ast+}}
             \else$\mathrm{D}^{\ast+}$\fi}
 \def\PDst{\ifmmode{\mathrm{D}^{\ast}}
             \else$\mathrm{D}^{\ast}$\fi}
 \def\PDstz{\ifmmode{\mathrm{D}^{\ast}(2010)^0}
            \else$\mathrm{D}^{\ast}(2010)^0$\fi}
 \def\PDiz{\ifmmode{\mathrm{D}_{1}(2420)^0}
           \else$\mathrm{D}_{1}(2420)^0$\fi}
 \def\PDstiiz{\ifmmode{\mathrm{D}^{\ast}_{2}(2460)^0}
              \else$\mathrm{D}^{\ast}_{2}(2460)^0$\fi}
 \def\PsDp{\ifmmode\mathrm{D_{s}^+}
           \else$\mathrm{D_{s}^+}$\fi}
 \def\PsDm{\ifmmode\mathrm{D_{s}^-}
           \else$\mathrm{D_{s}^-}$\fi}
 \def\PsDpm{\ifmmode\mathrm{D_{s}^{\pm}}
           \else$\mathrm{D_{s}^{\pm}}$\fi}
 \def\PsDst{\ifmmode\mathrm{D_{s}^{\ast}}
            \else$\mathrm{D_{s}^{\ast}}$\fi}
 \def\PsDipm{\ifmmode\mathrm{D_{s1}(2536)^{\pm}}
           \else$\mathrm{D_{s1}(2536)^{\pm}}$\fi}
 \def\PB{\ifmmode{\mathrm{B}}
          \else$\mathrm{B}$\fi}
 \def\PBp{\ifmmode{\mathrm{B}^{+}}
           \else$\mathrm{B}^{+}$\fi}
 \def\PBm{\ifmmode{\mathrm{B}^{-}}
           \else$\mathrm{B}^{-}$\fi}
 \def\PBpm{\ifmmode{\mathrm{B}^{\pm}}
            \else$\mathrm{B}^{\pm}$\fi}
 \def\PBz{\ifmmode{\mathrm{B}^0}
           \else$\mathrm{B}^0$\fi}
 \def\PbgL{\ifmmode{\mathrm{\Lambda}_b}
           \else$\mathrm{\Lambda}_b$\fi}
 \def\Pcgh{\ifmmode\mathrm{{\eta}_{c}(1S)}
           \else$\mathrm{{\eta}_{c}(1S)}$\fi}
 \def\PJgyy{\ifmmode\mathrm{J /\psi}
           \else$\mathrm{J /\psi}$\fi}
 \def\PJgy{\ifmmode\mathrm{J /\psi(1S)}
           \else$\mathrm{J /\psi(1S)}$\fi}
 \def\Pcgcz{\ifmmode\mathrm{{\chi}_{c0}(1P)}
            \else$\mathrm{{\chi}_{c0}(1P)}$\fi}
 \def\Pcgci{\ifmmode\mathrm{{\chi}_{c1}(1P)}
            \else$\mathrm{{\chi}_{c1}(1P)}$\fi}
 \def\Pcgcii{\ifmmode\mathrm{{\chi}_{c2}(1P)}
             \else$\mathrm{{\chi}_{c2}(1P)}$\fi}
 \def\Pgy{\ifmmode\mathrm{\psi(2S)}
          \else$\mathrm{\psi(2S)}$\fi}
 \def\Pgya{\ifmmode\mathrm{\psi(3770)}
           \else$\mathrm{\psi(3770)}$\fi}
 \def\Pgyb{\ifmmode\mathrm{\psi(4040)}
           \else$\mathrm{\psi(4040)}$\fi}
 \def\Pgyc{\ifmmode\mathrm{\psi(4160)}
           \else$\mathrm{\psi(4160)}$\fi}
 \def\Pgyd{\ifmmode\mathrm{\psi(4415)}
           \else$\mathrm{\psi(4415)}$\fi}
 \def\PgU{\ifmmode\mathrm{\Upsilon(1S)}
          \else$\mathrm{\Upsilon(1S)}$\fi}
 \def\Pbgcz{\ifmmode\mathrm{{\chi}_{b0}(1P)}
            \else$\mathrm{{\chi}_{b0}(1P)}$\fi}
 \def\Pbgci{\ifmmode\mathrm{{\chi}_{b1}(1P)}
            \else$\mathrm{{\chi}_{b1}(1P)}$\fi}
 \def\Pbgcii{\ifmmode\mathrm{{\chi}_{b2}(1P)}
             \else$\mathrm{{\chi}_{b2}(1P)}$\fi}
 \def\PgUa{\ifmmode\mathrm{\Upsilon(2S)}
           \else$\mathrm{\Upsilon(2S)}$\fi}
 \def\Pbgcza{\ifmmode\mathrm{{\chi}_{b0}(2P)}
             \else$\mathrm{{\chi}_{b0}(2P)}$\fi}
 \def\Pbgcia{\ifmmode\mathrm{{\chi}_{b1}(2P)}
             \else$\mathrm{{\chi}_{b1}(2P)}$\fi}
 \def\Pbgciia{\ifmmode\mathrm{{\chi}_{b2}(2P)}
              \else$\mathrm{{\chi}_{b2}(2P)}$\fi}
 \def\PgUb{\ifmmode\mathrm{\Upsilon(3S)}
           \else$\mathrm{\Upsilon(3S)}$\fi}
 \def\PgUc{\ifmmode\mathrm{\Upsilon(4S)}
           \else$\mathrm{\Upsilon(4S)}$\fi}
 \def\PgUd{\ifmmode\mathrm{\Upsilon(10860)}
           \else$\mathrm{\Upsilon(10860)}$\fi}
 \def\PgUe{\ifmmode\mathrm{\Upsilon(11020)}
           \else$\mathrm{\Upsilon(11020)}$\fi}
 \def\Pp{\ifmmode\mathrm{p}
         \else$\mathrm{p}$\fi}
 \def\Pap{\ifmmode\mathrm{\overline{p}}
         \else$\mathrm{\overline{p}}$\fi}
 \def\Pn{\ifmmode\mathrm{n}
         \else$\mathrm{n}$\fi}
 \def\PNa{\ifmmode\mathrm{N(1440)P_{11}}
          \else$\mathrm{N(1440)P_{11}}$\fi}
 \def\PNb{\ifmmode\mathrm{N(1520)D_{13}}
          \else$\mathrm{N(1520)D_{13}}$\fi}
 \def\PNc{\ifmmode\mathrm{N(1535)S_{11}}
          \else$\mathrm{N(1535)S_{11}}$\fi}
 \def\PNd{\ifmmode\mathrm{N(1650)S_{11}}
          \else$\mathrm{N(1650)S_{11}}$\fi}
 \def\PNe{\ifmmode\mathrm{N(1675)D_{15}}
          \else$\mathrm{N(1675)D_{15}}$\fi}
 \def\PNf{\ifmmode\mathrm{N(1680)F_{15}}
          \else$\mathrm{N(1680)F_{15}}$\fi}
 \def\PNg{\ifmmode\mathrm{N(1700)D_{13}}
          \else$\mathrm{N(1700)D_{13}}$\fi}
 \def\PNh{\ifmmode\mathrm{N(1710)P_{11}}
          \else$\mathrm{N(1710)P_{11}}$\fi}
 \def\PNi{\ifmmode\mathrm{N(1720)P_{13}}
          \else$\mathrm{N(1720)P_{13}}$\fi}
 \def\PNj{\ifmmode\mathrm{N(2190)G_{17}}
          \else$\mathrm{N(2190)G_{17}}$\fi}
 \def\PNk{\ifmmode\mathrm{N(2220)H_{19}}
          \else$\mathrm{N(2220)H_{19}}$\fi}
 \def\PNl{\ifmmode\mathrm{N(2250)G_{19}}
          \else$\mathrm{N(2250)G_{19}}$\fi}
 \def\PNm{\ifmmode\mathrm{N(2600)I_{1,11}}
          \else$\mathrm{N(2600)I_{1,11}}$\fi}
 \def\PgDa{\ifmmode\mathrm{\Delta(1232)P_{33}}
           \else$\mathrm{\Delta(1232)P_{33}}$\fi}
 \def\PgDb{\ifmmode\mathrm{\Delta(1620)S_{31}}
           \else$\mathrm{\Delta(1620)S_{31}}$\fi}
 \def\PgDc{\ifmmode\mathrm{\Delta(1700)D_{33}}
           \else$\mathrm{\Delta(1700)D_{33}}$\fi}
 \def\PgDd{\ifmmode\mathrm{\Delta(1900)S_{31}}
           \else$\mathrm{\Delta(1900)S_{31}}$\fi}
 \def\PgDe{\ifmmode\mathrm{\Delta(1905)F_{35}}
           \else$\mathrm{\Delta(1905)F_{35}}$\fi}
 \def\PgDf{\ifmmode\mathrm{\Delta(1910)P_{31}}
           \else$\mathrm{\Delta(1910)P_{31}}$\fi}
 \def\PgDh{\ifmmode\mathrm{\Delta(1920)P_{33}}
           \else$\mathrm{\Delta(1920)P_{33}}$\fi}
 \def\PgDi{\ifmmode\mathrm{\Delta(1930)D_{35}}
           \else$\mathrm{\Delta(1930)D_{35}}$\fi}
 \def\PgDj{\ifmmode\mathrm{\Delta(1950)F_{37}}
           \else$\mathrm{\Delta(1950)F_{37}}$\fi}
 \def\PgDk{\ifmmode\mathrm{\Delta(2420)H_{3,11}}
           \else$\mathrm{\Delta(2420)H_{3,11}}$\fi}
 \def\PgDpp{\ifmmode\mathrm{\Delta^{++}}
           \else$\mathrm{\Delta^{++}}$\fi}
 \def\PgL{\ifmmode\mathrm{\Lambda}
          \else$\mathrm{\Lambda}$\fi}
 \def\PagL{\ifmmode\mathrm{\overline{\Lambda}}
            \else$\mathrm{\overline{\Lambda}}$\fi}
 \def\PgLa{\ifmmode\mathrm{\Lambda(1405) S_{01}}
           \else$\mathrm{\Lambda(1405) S_{01}}$\fi}
 \def\PgLb{\ifmmode\mathrm{\Lambda(1520) D_{03}}
           \else$\mathrm{\Lambda(1520) D_{03}}$\fi}
 \def\PgLc{\ifmmode\mathrm{\Lambda(1600) P_{01}}
           \else$\mathrm{\Lambda(1600) P_{01}}$\fi}
 \def\PgLd{\ifmmode\mathrm{\Lambda(1670) S_{01}}
           \else$\mathrm{\Lambda(1670) S_{01}}$\fi}
 \def\PgLe{\ifmmode\mathrm{\Lambda(1690) D_{03}}
           \else$\mathrm{\Lambda(1690) D_{03}}$\fi}
 \def\PgLf{\ifmmode\mathrm{\Lambda(1800) S_{01}}
           \else$\mathrm{\Lambda(1800) S_{01}}$\fi}
 \def\PgLg{\ifmmode\mathrm{\Lambda(1810) P_{01}}
           \else$\mathrm{\Lambda(1810) P_{01}}$\fi}
 \def\PgLh{\ifmmode\mathrm{\Lambda(1820) F_{05}}
           \else$\mathrm{\Lambda(1820) F_{05}}$\fi}
 \def\PgLi{\ifmmode\mathrm{\Lambda(1830) D_{05}}
           \else$\mathrm{\Lambda(1830) D_{05}}$\fi}
 \def\PgLj{\ifmmode\mathrm{\Lambda(1890) P_{03}}
           \else$\mathrm{\Lambda(1890) P_{03}}$\fi}
 \def\PgLk{\ifmmode\mathrm{\Lambda(2100) G_{07}}
           \else$\mathrm{\Lambda(2100) G_{07}}$\fi}
 \def\PgLl{\ifmmode\mathrm{\Lambda(2110) F_{05}}
           \else$\mathrm{\Lambda(2110) F_{05}}$\fi}
 \def\PgLm{\ifmmode\mathrm{\Lambda(2350) H_{09}}
           \else$\mathrm{\Lambda(2350) H_{09}}$\fi}
 \def\PgS{\ifmmode{\rm \Sigma}
           \else${\rm \Sigma}$\fi}
 \def\PgSp{\ifmmode\mathrm{\Sigma^+}
           \else$\mathrm{\Sigma^+}$\fi}
 \def\PgSz{\ifmmode\mathrm{\Sigma^0}
           \else$\mathrm{\Sigma^0}$\fi}
 \def\PgSm{\ifmmode\mathrm{\Sigma^-}
           \else$\mathrm{\Sigma^-}$\fi}
 \def\PgSpm{\ifmmode\mathrm{\Sigma^{\pm}}
           \else$\mathrm{\Sigma^{\pm}}$\fi}
 \def\PgSa{\ifmmode\mathrm{\Sigma(1385) P_{13}}
           \else$\mathrm{\Sigma(1385) P_{13}}$\fi}
 \def\PgSb{\ifmmode\mathrm{\Sigma(1660) P_{11}}
           \else$\mathrm{\Sigma(1660) P_{11}}$\fi}
 \def\PgSc{\ifmmode\mathrm{\Sigma(1670) D_{13}}
           \else$\mathrm{\Sigma(1670) D_{13}}$\fi}
 \def\PgSd{\ifmmode\mathrm{\Sigma(1750) S_{11}}
           \else$\mathrm{\Sigma(1750) S_{11}}$\fi}
 \def\PgSe{\ifmmode\mathrm{\Sigma(1775) D_{15}}
           \else$\mathrm{\Sigma(1775) D_{15}}$\fi}
 \def\PgSf{\ifmmode\mathrm{\Sigma(1915) F_{15}}
           \else$\mathrm{\Sigma(1915) F_{15}}$\fi}
 \def\PgSg{\ifmmode\mathrm{\Sigma(1940) D_{13}}
           \else$\mathrm{\Sigma(1940) D_{13}}$\fi}
 \def\PgSh{\ifmmode\mathrm{\Sigma(2030) F_{17}}
           \else$\mathrm{\Sigma(2030) F_{17}}$\fi}
 \def\PgSi{\ifmmode\mathrm{\Sigma(2050)}
           \else$\mathrm{\Sigma(2050)}$\fi}
 \def\PgXz{\ifmmode\mathrm{\Xi^0}
           \else$\mathrm{\Xi^0}$\fi}
 \def\PgXm{\ifmmode\mathrm{\Xi^-}
           \else$\mathrm{\Xi^-}$\fi}
 \def\PgXa{\ifmmode\mathrm{\Xi(1530)}
           \else$\mathrm{\Xi(1530)}$\fi}
 \def\PgXas{\ifmmode\mathrm{\Xi(1530)P_{13}}
           \else$\mathrm{\Xi(1530)P_{13}}$\fi}
 \def\PgXb{\ifmmode\mathrm{\Xi(1690)}
           \else$\mathrm{\Xi(1690)}$\fi}
 \def\PgXbb{\ifmmode\mathrm{\Xi(1620)}
           \else$\mathrm{\Xi(1620)}$\fi}
 \def\PgXc{\ifmmode\mathrm{\Xi(1820)D_{13}}
           \else$\mathrm{\Xi(1820)D_{13}}$\fi}
 \def\PgXcs{\ifmmode\mathrm{\Xi(1820)}
           \else$\mathrm{\Xi(1820)}$\fi}
 \def\PgXd{\ifmmode\mathrm{\Xi(1950)}
           \else$\mathrm{\Xi(1950)}$\fi}
 \def\PgXe{\ifmmode\mathrm{\Xi(2030)}
           \else$\mathrm{\Xi(2030)}$\fi}
 \def\PgOm{\ifmmode\mathrm{\Omega^-}
           \else$\mathrm{\Omega^-}$\fi}
 \def\PgO{\ifmmode\mathrm{\Omega}
           \else$\mathrm{\Omega}$\fi}
 \def\PgOma{\ifmmode\mathrm{\Omega(2250)^-}
            \else$\mathrm{\Omega(2250)^-}$\fi}
 \def\PcgL{\ifmmode\mathrm{\Lambda_c}
            \else$\mathrm{\Lambda_c}$\fi}
 \def\PacgL{\ifmmode\mathrm{\overline{\Lambda}_c}
            \else$\mathrm{\overline{\Lambda}_c}$\fi}
 \def\PcgLp{\ifmmode\mathrm{\Lambda_c^+}
            \else$\mathrm{\Lambda_c^+}$\fi}
 \def\PcgLm{\ifmmode{\rm \Lambda_c^-}
            \else${\rm \Lambda_c^-}$\fi}
 \def\PcgX{\ifmmode\mathrm{\Xi_c}
            \else$\mathrm{\Xi_c}$\fi}
 \def\PcgXz{\ifmmode\mathrm{\Xi_c^0}
            \else$\mathrm{\Xi_c^0}$\fi}
 \def\PcgXp{\ifmmode\mathrm{\Xi_c^+}
            \else$\mathrm{\Xi_c^+}$\fi}
 \def\PcgS{\ifmmode\mathrm{\Sigma_c}
           \else$\mathrm{\Sigma_c}$\fi}
 \def\PcgSz{\ifmmode\mathrm{\Sigma_c^0}
           \else$\mathrm{\Sigma_c^0}$\fi}
 \def\PcgSp{\ifmmode\mathrm{\Sigma_c^+}
           \else$\mathrm{\Sigma_c^+}$\fi}
 \def\PcgSpp{\ifmmode\mathrm{\Sigma_c^{++}}
           \else$\mathrm{\Sigma_c^{++}}$\fi}
 \def\PcgO{\ifmmode{\mathrm \Omega_c}
           \else${\mathrm \Omega_c}$\fi}
 \def\PcgOz{\ifmmode{\mathrm \Omega_c^{0}}
           \else${\mathrm \Omega_c^{0}}$\fi}
 \def\PSgg{\ifmmode\mathrm{\tilde{\gamma}}
           \else$\mathrm{\tilde{\gamma}}$\fi}
 \def\PSgxz{\ifmmode\mathrm{\tilde{\chi}^0_i}
            \else$\mathrm{\tilde{\chi}^0_i}$\fi}
 \def\PSZz{\ifmmode\mathrm{\tilde{Z}^0}
           \else$\mathrm{\tilde{Z}^0}$\fi}
 \def\PSHz{\ifmmode\mathrm{\tilde{H}^0_j}
           \else$\mathrm{\tilde{H}^0_j}$\fi}
 \def\PSgxpm{\ifmmode\mathrm{\tilde{\chi}^{\pm_i}}
             \else$\mathrm{\tilde{\chi}^{\pm_i}}$\fi}
 \def\PSWpm{\ifmmode\mathrm{\tilde{W}^{\pm}}
            \else$\mathrm{\tilde{W}^{\pm}}$\fi}
 \def\PSHpm{\ifmmode\mathrm{\tilde{H}^{\pm_j}}
            \else$\mathrm{\tilde{H}^{\pm_j}}$\fi}
 \def\PSgn{\ifmmode\mathrm{\tilde{\nu}}
           \else$\mathrm{\tilde{\nu}}$\fi}
 \def\PSe{\ifmmode\mathrm{\tilde{e}}
          \else$\mathrm{\tilde{e}}$\fi}
 \def\PSgm{\ifmmode\mathrm{\tilde{\mu}}
           \else$\mathrm{\tilde{\mu}}$\fi}
 \def\PSgt{\ifmmode\mathrm{\tilde{\tau}}
           \else$\mathrm{\tilde{\tau}}$\fi}
 \def\PSq{\ifmmode\mathrm{\tilde{q}}
          \else$\mathrm{\tilde{q}}$\fi}
 \def\PSg{\ifmmode\mathrm{\tilde{g}}
          \else$\mathrm{\tilde{g}}$\fi}

\def\insta {University of Bristol, Bristol, United Kingdom}
\def\instb {CERN, CH-1211 Gen\`eve 23, Switzerland}
\def\instc {Genoa University/INFN, Dipartimento di Fisica,I-16146 Genova, Italy}
\def\instd {Grenoble ISN, F-38026 Grenoble, France}
\def\inste {Max-Planck-Institut f\"ur Kernphysik, Postfach 103980, D-69029 Heidelberg, Germany}
\def\instf {Universit\"at Heidelberg, Physikalisches Institut, D-69120 Heidelberg, Germany}
\def\instg {Universit\"at Mainz, Institut f\"ur Kernphysik, D-55099 Mainz, Germany}
\def\insth {Moscow Lebedev Physics Institute, RU-117924, Moscow, Russia}
\def\insti {University of Iowa, Iowa City, IA 52242, USA}
\def\instj {Rutgers University, Piscataway, New Jersey 08854, USA}
\def\instk {NIKHEF, 1009 D8 Amsterdam, The Netherlands}

\begin{document}

\preprint{xxx}

\title[Search for Ximinusminus]
{Search for the exotic $\Xi^{--}(1860)$ Resonance in 340\gevc1\
$\PgSm$-Nucleus Interactions}

\author{M.I.~Adamovich}
\thanks{Deceased.}
\affiliation{\insth}
\author{Yu.A.~Alexandrov}
\thanks{Supported by the Deutsche Forschungsgemeinschaft,
           contract number436 RUS 113/465, and Russian Foundation for
           Basic Research under contract number RFFI 98-02-04096.}
\affiliation{\insth}
\author{S.P.~Baranov}
\affiliation{\insth}
\author{D.~Barberis}
\affiliation{\instc}
\author{M.~Beck}
\affiliation{\inste}
\author{C.~B\'erat}
\affiliation{\instd}
\author{W.~Beusch}
\affiliation{\instb}
\author{M.~Boss}
\affiliation{\instf}
\thanks{Supported by the Bundesministerium f\"ur Bildung, Wissenschaft,
Forschung und Technologie, Germany, under contract numbers
05HD515I and 06HD524I.}
\author{S.~Brons}
\altaffiliation{Present address: TRIUMF, Vancouver, B.C., Canada
V6T 2A3} \affiliation{\inste}
\author{W.~Br\"uckner}
\affiliation{\inste}
\author{M.~Bu\'enerd}
\affiliation{\instd}
\author{C.~Busch}
\affiliation{\instf}
\author{C.~B\"uscher}
\affiliation{\inste}
\author{F.~Charignon}
\affiliation{\instd}
\author{J.~Chauvin}
\affiliation{\instd}
\author{E.A.~Chudakov}
\altaffiliation{Present address: Thomas Jefferson Lab, Newport
News, VA 23606, USA.} \affiliation{\instf}
\author{U.~Dersch}
\affiliation{\inste}
\author{F.~Dropmann}
\affiliation{\inste}
\author{J.~Engelfried}
\altaffiliation{Present address: Institudo de Fisica, Universidad
San Luis Potosi, S.L.P. 78240, Mexico.} \affiliation{\instf}
\author{F.~Faller}
\altaffiliation{Present address: Fraunhofer Institut f\"ur
Solarenergiesysteme, D-79100 Freiburg, Germany.}
\affiliation{\instf}
\author{A.~Fournier}
\affiliation{\instd}
\author{S.G.~Gerassimov}
\altaffiliation{Present address: Fakult\"at f\"ur Physik,
Universit\"at Freiburg, Germany.} \affiliation{\inste}
\affiliation{\insth}
\author{M.~Godbersen}
\affiliation{\inste}
\author{P.~Grafstr\"om}
\affiliation{\instb}
\author{Th.~Haller}
\affiliation{\inste}
\author{M.~Heidrich}
\affiliation{\inste}
\author{E.~Hubbard}
\affiliation{\inste}
\author{R.B.~Hurst}
\affiliation{\instc}
\author{K.~K\"onigsmann}
\altaffiliation{Present address: Fakult\"at f\"ur Physik,
Universit\"at Freiburg, Germany.} \affiliation{\inste}
\author{I.~Konorov}
\altaffiliation{Present address: Technische Universit\"at
M\"unchen, Garching, Germany.} \affiliation{\inste}
\affiliation{\insth}
\author{N.~Keller}
\affiliation{\instf}
\author{K.~Martens}
\altaffiliation{Present address: Department of Physics and
Astronomy, SUNY at Stony Brook, NY 11794-3800, USA.}
\affiliation{\instf}
\author{Ph.~Martin}
\affiliation{\instd}
\author{S.~Masciocchi}
\altaffiliation{Present address: Max-Planck-Institut f\"ur Physik,
M\"unchen, Germany.} \affiliation{\inste}
\author{R.~Michaels}
\altaffiliation{Present address: Thomas Jefferson Lab, Newport
News, VA 23606, USA.} \affiliation{\inste}
\author{U.~M\"uller}
\affiliation{\instg}
\thanks{Supported by the Bundesministerium f\"ur Bildung, Wissenschaft,
Forschung und Technologie, Germany, under contract number 06MZ5265.}
\author{H.~Neeb}
\affiliation{\inste}
\author{D.~Newbold}
\affiliation{\insta}
\author{C.~Newsom}
\affiliation{\insti}
\author{S.~Paul}
\altaffiliation{Present address: Technische Universit\"at
M\"unchen, Garching, Germany.} \affiliation{\inste}
\author{J.~Pochodzalla}
\email[Contact person:]{pochodza@kph.uni-mainz.de}
\altaffiliation{present address: Universit\"at Mainz, Institut
f\"ur Kernphysik, D-55099 Mainz, Germany.} \affiliation{\inste}
\author{I.~Potashnikova}
\affiliation{\inste}
\author{B.~Povh}
\affiliation{\inste}
\author{R.~Ransome}
\affiliation{\instj}
\author{Z.~Ren}
\affiliation{\inste}
\author{M.~Rey-Campagnolle}
\altaffiliation {Present address: CERN, CH-1211 Gen\`eve 23,
Switzerland} \affiliation{\instd}
\author{G.~Rosner}
\altaffiliation {Present address: Dept. of Physics and Astronomy,
University of Glasgow, Glasgow G12 8QQ, United Kingdom}
\affiliation{\instg}
\author{L.~Rossi}
\affiliation{\instc}
\author{H.~Rudolph}
\affiliation{\instg}
\author{C.~Scheel}
\affiliation{\instk}
\author{L.~Schmitt}
\altaffiliation{Present address: Technische Universit\"at
M\"unchen, Garching, Germany.} \affiliation{\instg}
\author{H.-W.~Siebert}
\altaffiliation{Present address: Universit\"at Mainz, Institut
f\"ur Kernphysik, D-55099 Mainz, Germany.}\affiliation{\instf}
\author{A.~Simon}
\altaffiliation{Present address: Fakult\"at f\"ur Physik,
Universit\"at
 Freiburg, Germany.}
\affiliation{\instf}
\author{V.J.~Smith}
\altaffiliation{Supported by the UK PPARC} \noaffiliation
\affiliation{\insta}
\author{O.~Thilmann}
\affiliation{\instf}
\author{A.~Trombini}
\affiliation{\inste}
\author{E.~Vesin}
\affiliation{\instd}
\author{B.~Volkemer}
\affiliation{\instg}
\author{K.~Vorwalter}
\affiliation{\inste}
\author{Th.~Walcher}
\affiliation{\instg}
\author{G.~W\"alder}
\affiliation{\instf}
\author{R.~Werding}
\affiliation{\inste}
\author{E.~Wittmann}
\affiliation{\inste}
\author{M.V.~Zavertyaev}
\altaffiliation{Supported by the Deutsche Forschungsgemeinschaft,
contract number436 RUS 113/465, and Russian Foundation for Basic
Research under contract number RFFI 98-02-04096.}
\affiliation{\insth}

\collaboration{WA89 collaboration} \noaffiliation

\date{\today}

\begin{abstract}
We report on a high statistics search for the $\Xi^{--}(1860)$
resonance in $\Sigma^-$-nucleus collisions at 340{\gevc1}. No
evidence for this resonance is found in our data sample which
contains 676000 $\Xi^-$ candidates above background. For the decay
channel $\Xi^{--}(1860) \rightarrow \Xi^-\pi^-$ and the kinematic
range 0.15$<x_F<$0.9 we find a 3$\sigma$ upper limit for the
production cross section of 3.1 and 3.5 $\mu$b per nucleon for
reactions with carbon and copper, respectively.
\end{abstract}

\pacs{13.85.-t,13.85.Rm,25.80.Pw}

\maketitle

At present eleven experimental groups have reported evidence for a
narrow baryonic resonance in the KN channel  at a mass of about
1530\Mevc2
\cite{Theta:LEPS,Theta:DIANA,Theta:CLASd,Theta:SAPHIR,Theta:CLASp,
Theta:NEUTRINO,Theta:HERMES,Theta:SVD,Theta:COSY,Theta:bubble,Theta:ZEUS}
(for an updated list of references see \cite{Theta:Lit}). Based on
previous predictions \cite{Theta:Diakonov} (for some earlier
references see also \cite{Theta:Walliser}) this resonance was
interpreted as a pentaquark state. However, doubts have been
raised because of possible experimental artefacts
\cite{Theta:Dzierba,Theta:Zavertyaev} and, furthermore,
interpretations in terms of more conventional processes are under
discussion
\cite{Theta:Bicudo,Theta:Nussinov,Theta:Kahana,Theta:Kishimoto}
(see however Ref. \cite{Theta:Llanes}). A common drawback of the
individual observations is the limited statistics and hence
limited confidence \cite{Bit00} of the peaks.

The interpretation of the observed peaks in terms of a five-quark
state was significantly strengthened by the subsequent observation
of another member of the anti\-cipated antidecuplet of
pentaquarks. Based on 1640 $\Xi^-$ candidates produced in p+p
interactions at 160{\gevc1} beam momentum, both in the
$\Xi^-\pi^+$ and the $\Xi^-\pi^-$ channels narrow peak structures
at an invariant mass of 1.860{\Gevc2} were observed by the NA49
collaboration \cite{Xi:NA49}. Possible signals of a  $\Xi^*$
resonance at 1.860{\Gevc2} decaying into ${\Xi^-}\pi^+$ and
$Y\overline{K}$ were reported already 1977 for K$^-$p interactions
at 2.87{\gevc1} \cite{Bri77}. However, no corresponding signals
have been seen in other K$^-$ induced reactions (for a compilation
and a discussion of these data see Ref. \cite{Xi:Fischer}). A
preliminary analysis of proton-nucleus interactions at 920\gevc1
by the HERA-B collaboration using a total of 19000 reconstructed
$\Xi^-$ and $\overline{\Xi}^+$ events, shows no indication for the
$\Xi^{--}$ nor the $\Theta^+$ resonances \cite{Xi:HERAB}. Searches
for the $\Xi(1860)$ resonances are also being performed by the
ZEUS and the CDF collaboration. The ZEUS data comprise 1361
$\Xi^-$ and 1303 $\overline{\Xi}^+$ events, the CDF sample
contains 19150 $\Xi^-$ and 16736 $\overline{\Xi}^+$. Negative --
though still preliminary -- results have been reported at the
DIS04 conference \cite{Xi:CDFZEUS}.

\begin{figure}[tb]
\includegraphics[width=8.cm]{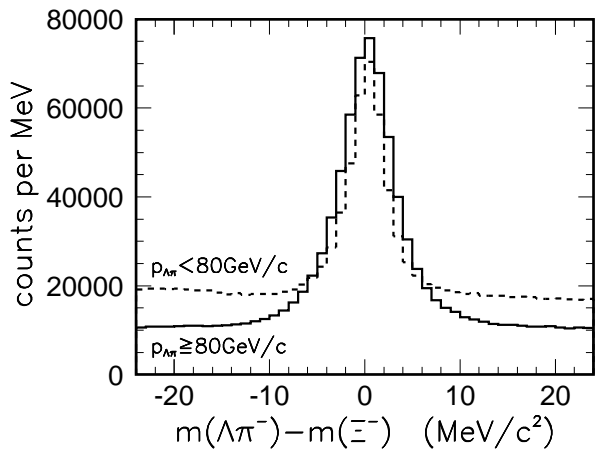}
\vspace{-0.3cm} \caption{Invariant mass distributions of
$\Lambda\pi^-$ pairs with p$_{\Lambda\pi} \geq$ 80{\gevc1} (solid
histogram) and $< $80{\gevc1} (dashed histogram) in 340{\gevc1}
$\Sigma^-$ induced interactions.} \label{fig:wa8901}
\end{figure}

It is indisputable that further high-statistics experiments are
needed to establish the observed resonances beyond any doubt and
to determine the quantum numbers of these states if they exist.
Moreover, the observation (or non-observation) of these resonances
in different reactions may help to shed some light on the
production mechanism and possibly also on the internal structure
of these exotic states.

The hyperon beam experiment WA89 had the primary goal to study
charmed particles and their decays. At the same time it collected
a high statistics data sample of hyperons and hyperon resonances
\cite{{WA89:xi},{WA89:ximpip},{WA89:xistar},{WA89:sigma},{WA89:ll},{WA89:v0},{WA89:xipol}}.
Here we present a search for the S=-2 resonance in $\Sigma^-$
induced reactions on C and Cu at 340 GeV/c . We also include
interactions in the tracking detectors (silicon detectors and
plastic scintillator) located close to these targets.

The hyperon beamline \cite{WA89:beam} selected $\Sigma^-$ hyperons
with a mean momentum of 340~\gevc1 and a momentum spread of
$\sigma (p)/p=9\%$. Although the actual $\pi^-$ to $\Sigma^-$
ratio of the beam was about 2.3, high-momentum pions were strongly
suppressed at the trigger level by a set of transition radiation
detectors \cite{WA89:TRD} resulting in a remaining pion
contamination of about 12\%. In addition the beam contained small
admixtures of $K^-$ (2.1\%) and $\Xi^-$ (1.3\%) \cite{WA89:xi}.
The trajectories of incoming and outgoing particles were measured
in silicon microstrip detectors upstream and downstream of the
target. The experimental target itself consisted of one copper
slab with a thickness of 0.025 $\lambda_I$ in beam direction,
followed by three carbon (diamond powder) slabs of 0.008
$\lambda_I$ each, where $\lambda_I$ is the interaction length.

\begin{figure}[t]
\includegraphics[width=8.cm]{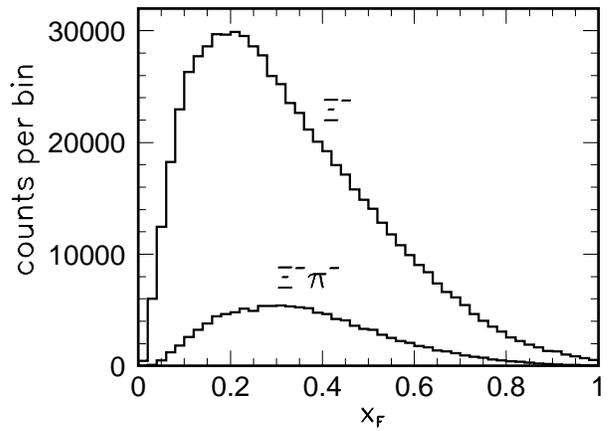}
\vspace{-0.3cm} \caption{Upper histogram: $x_F$ distribution of
the observed $\Xi^-$ events within a $\pm$2$\sigma$ mass window.
Lower histogram:  $x_F$ distribution of the observed $\Xi^-\pi^-$
pairs within the mass range between 1.82 and 1.90 {\Gevc2}. In
both cases the background has been subtracted by means of sideband
events.} \label{fig:wa8902}
\end{figure}

The momenta of the decay particles were measured in a magnetic
spectrometer equipped with MWPCs and drift chambers. In order to
allow hyperons and $K^0_S$ emerging from the target to decay in
front of the magnet the target was placed 13.6m upstream of the
center of the spectrometer magnet. The apparatus also comprised a
ring-imaging Cherenkov detector, a lead glass electromagnetic
calorimeter and an lead/scintillator hadron calorimeter, which
were not used in this analysis.

$\Xi^-$ were reconstructed in the decay chain $\Xi^- \rightarrow
\Lambda\pi^- \rightarrow p\pi^-\pi^-$. The invariant mass
distributions of the $\Xi^-$ candidates are shown Fig.
\ref{fig:wa8901} for two regions of the total momentum of the
$\Lambda\pi$ pair. The cut at 80{\gevc1} corresponds to an $x_F$
value of about 0.25 (see below). In our data sample the central
peak-to-background ratio varies between about 4 at small momenta
and 8 at larger momenta. The rms-width of the $\Xi^-$ peak can be
approximated by the relation
$\sigma=\sqrt{3.5{MeV^2/c^4}+2.2\cdot10^{-10}p_{\Xi}^2/c^2}$ where
$p_{\Xi}$ denotes the total momentum of the $\Lambda\pi$ pair.
$\Xi^-$ candidates within a $\pm$2$\sigma$ window around the
nominal $\Xi^-$ mass were used in the further analysis. The
present analysis is based on a total of 676k $\Xi^-$ candidates
observed over a background of 170k $p\pi^-\pi^-$ combinations
\cite{WA89:events}. Out of these candidates 240k, 281k and 155k
can be attributed to the C, Cu and "Si+C+H" target, respectively.

Because of the strangeness content of the $\Sigma^-$ beam the
cross sections for $\Xi$ resonances are shifted towards large
$x_F$ with respect to the $\Sigma^-$-nucleon cm-system
\cite{WA89:xistar}. Since in the WA89 setup the efficiency drops
significantly at $x_F<$0.1 the yield of $\Xi^-$ peaks at $x_F
\approx$ 0.2 (upper histogram in Fig. \ref{fig:wa8902}).
$\Xi^-\pi^-$ pairs within the mass range of 1.82 to 1.90 {\Gevc2}
are shifted to even larger $x_F$ (lower histogram in
Fig.~\ref{fig:wa8902}). In both cases background was subtracted by
means of two 2$\sigma$ wide sidebands located at
[-24{\Mevc2},-24{\Mevc2}+2$\sigma$] and
[24{\Mevc2}-2$\sigma$,24{\Mevc2}] (cf. Fig. \ref{fig:wa8901}). For
comparison, the $\Xi^-$ events observed by NA49 are distributed
over an $x_F$ range between -0.25 and +0.25 \cite{NA49:Barna}.
\begin{figure}[t]
\includegraphics[width=8.5cm]{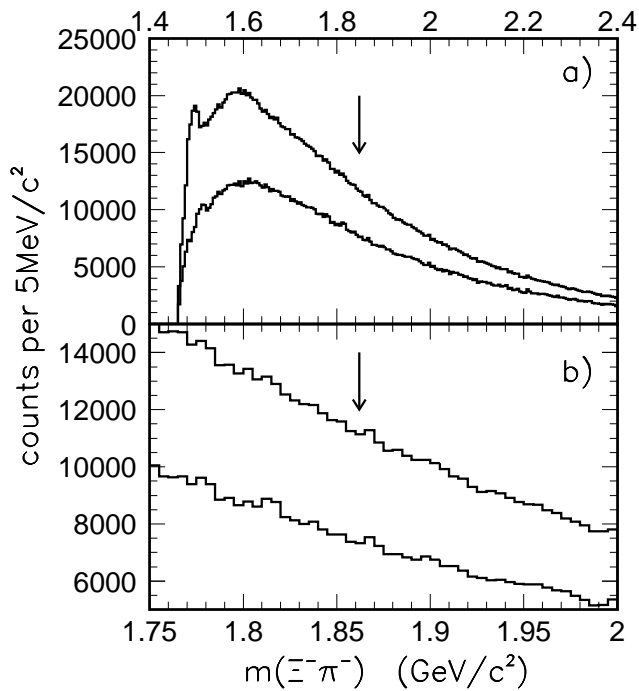}
\vspace{-0.3cm} \caption{Effective mass distribution of
$\Xi^-\pi^-$ combinations of all reactions, including also
reactions in the tracking detectors (Si+C+H) close to the C and Cu
targets.  Part b) shows an extended view of the region around
1.862\Gevc2\ marked by the arrows. Note the offset of the y-axis
in this panel. In each panel the lower histogram shows the
distribution after background subtraction via sidebands.}
\label{fig:wa8903}
\end{figure}

Fig. \ref{fig:wa8903} shows the invariant mass spectrum of all
observed $\Xi^-\pi^-$ pairs. Fig. \ref{fig:wa8903}b shows an
extended view of the region around a mass of 1.862\Gevc2\ marked
by the arrows. All reactions, including also interactions in the
tracking detectors close to the C and Cu targets, contribute to
this figure. The structure observed at around 1.5\Gevc2 in the
upper histogram of Fig.~\ref{fig:wa8903}a is caused by events
where the negative pion from the decay of the $\Xi^-$ was wrongly
reconstructed as a double track. As can be seen from the lower
histogram in Fig. \ref{fig:wa8903}a, these fake pairs are reduced
substantially by subtracting background from $\Xi^-$ sideband
events.
\begin{figure}[t]
\includegraphics[width=8.5cm]{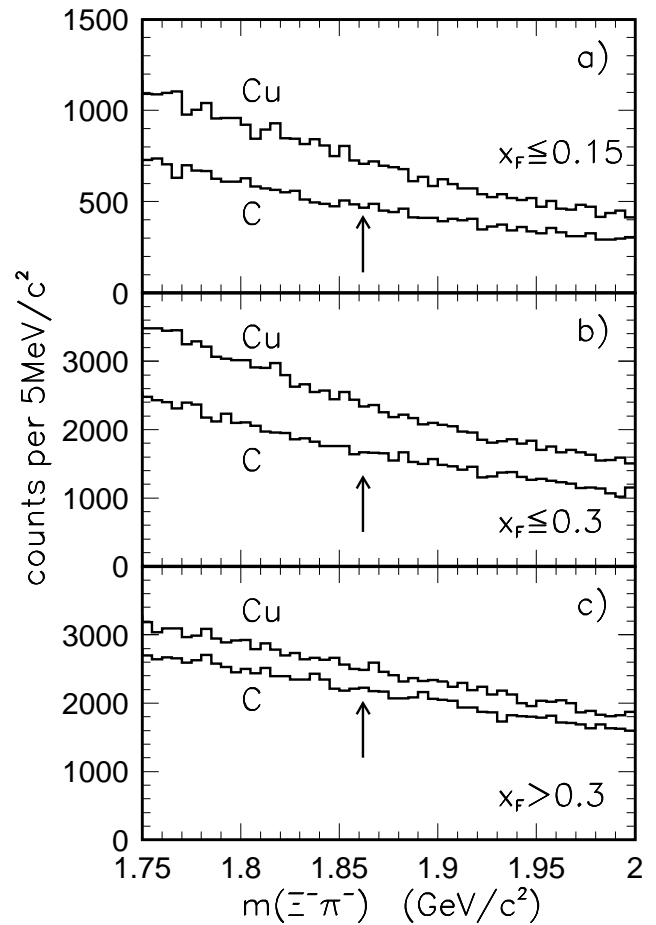}
\vspace{-0.3cm} \caption{Effective mass distribution of
$\Xi^-\pi^-$ combinations with $x_F(\Xi^-\pi^-)\leq$0.15 (part a),
$x_F(\Xi^-\pi^-)\leq$0.3 (part b) and $x_F(\Xi^-\pi^-) > $0.3
(part c). In each plot the lower and upper histogram correspond to
the carbon and copper target, respectively.} \label{fig:wa8904}
\end{figure}

The NA49 collaboration has observed a ratio of $\Xi^{--}$ to
$\Xi^-$ candidates of about 1/40. If we assume the same {\it
relative} production cross sections over the full kinematic range
for the reaction in question and similar {\it relative} detection
efficiencies
$[{\varepsilon}(\Xi^{--})/{\varepsilon}(\Xi^-)]_{WA89}\approx
[{\varepsilon}(\Xi^{--})/{\varepsilon}(\Xi^-)]_{NA49}$ we would
expect of the order of 17000 $\Xi^{--} \rightarrow \Xi^-+\pi^-$
events in our full data sample. The FWHM of the peaks observed by
NA49 is 17\Mevc2 and is limited by the experimental resolution.
Since in our experiment the resolution is expected to be slightly
smaller $\approx$ 10{\Mevc2} (FWHM), this excess should be
concentrated in less than 6 channels in Fig. \ref{fig:wa8903}b.
Obviously, no such enhancement can be seen in the spectra.

The $\Xi(1860)$ events observed by NA49 are concentrated at small
$x_F$. For a better comparison with the NA49 experiment we
therefore scanned our data for different ranges of $x_F$.
Fig.\ref{fig:wa8904} shows the effective mass distributions of
$\Xi^-\pi^-$ combinations with $x_F(\Xi^-\pi^-)\leq$0.15,
$\leq$0.3 and $> $0.3 in the region around 1.862{\Gevc2}. In each
panel, the upper and lower histograms correspond to reactions with
the carbon and copper target, respectively. No background
subtraction was applied to these spectra.  Assuming again a
$\Xi^{--}$ to $\Xi^-$ ratio of 1/40 as observed by NA49 and
considering now only the $x_F$ range between 0 and 0.15, we
estimate that approximately 700 and 900 $\Xi^{--} \rightarrow
\Xi^-\pi^-$ events should be seen in Fig.\ref{fig:wa8904}a for the
C and Cu target, respectively. None of these spectra shows
evidence for a statistically significant signal around
1.862{\Gevc2}, nor does such a signal appear in any other
sub-sample.

\begin{table}[t]
\caption{3$\sigma$ limits $n_{max}$ for the number of events and
the corresponding limits on the differential cross section for
$\Xi^{--}(1860)$ production in copper and carbon, for different
$x_F$ intervals. \label{tab:wa8901}}
\begin{ruledtabular}
\begin{tabular}{cccccc}
\multicolumn{3}{c}{ } & \multicolumn
{2}{c}{BR$\cdot d\sigma$/d$x_{F,max} $ [$\mu$b]}\\
target& $x_F$ & $n_{max}$ & per nucleus & per nucleon \\
\colrule

Cu& ~~~0.-0.15 & 170&-\footnotemark[1]&-\footnotemark[1]\\
  & 0.15-0.30 & 270&170&11\\
  & 0.30-0.45 & 300&190&12\\
  & 0.45-0.60 & 220&160&10\\
  & 0.60-0.75 & 180&150&9\\
  & 0.75-0.90 &  85&150&10\\ \hline
C & ~~~0.-0.15 & 140&-\footnotemark[1]&-\footnotemark[1]\\
  & 0.15-0.30 & 240& 62&12\\
  & 0.30-0.45 & 220& 50&10\\
  & 0.45-0.60 & 180& 52&10\\
  & 0.60-0.75 & 140& 46&9\\
  & 0.75-0.90 &  60& 28& 5\\
\end{tabular}
\end{ruledtabular}
\footnotetext[1]{The sharp rise of the efficiency between
$x_F$=0.05 and 0.15 which is reflected in the observed $x_F$
distributions (Fig. \ref{fig:wa8902}), prevents a reliable
determination of the cross section below $x_F\leq$0.15.}

\end{table}

Upper limits on the production cross sections were estimated
separately for the copper and carbon targets, in five bins of
$x_F$ between $x_F=0.15$ and $x_F=0.9$. For this purpose, we
calculated limits, $n_{max}$, on the number of $\Xi^{--} (1860)
\rightarrow \Xi^- +\pi^-$ decays as follows: Based on the claimed
experimental width of the $\Xi^{--} (1860)$ of $<17 \, MeV/c^2$
FWHM \cite{Xi:NA49}, we calculated  $n_{max}$ from the observed
number of $\Xi^- \pi^-$ combinations, $n_i$, inside three mass
windows of 20\Mevc2 width, centered at 1850, 1860 and 1870
{\Mevc2}, resp., for $i=1,2,3$. From a fit to the observed $\Xi^-
\pi^-$ mass spectrum between 1700 and 2000 $MeV/c^2$ (excluding
the presumed signal region), we calculated the non-resonant
backgrounds $b_i$ in each bin. The 3$\sigma$ limits were then
obtained by the formula $n_{max}\, =\, max_{i=1,2,3}
\{max(0,n_i-b_i)\, +\, 3\sqrt{b_i} \}$ and are listed in column 3
of Tab. \ref{tab:wa8901}. From these numbers we derived the upper
limits on the product of $BR$, the decay branching ratio, and the
differential production cross sections $d\sigma /dx_F$ {\em per
nucleus} given in column 4 of Tab. \ref{tab:wa8901}. Assuming a
dependence of the cross section on the mass number as
$\sigma_{nucl}\propto \sigma_0\cdot A^{2/3}$, where $\sigma_0$ is
the cross section {\em per nucleon}, we finally obtained the
limits on $BR\cdot d\sigma_0 /dx_F$ in the last column of the
table.

\begin{table}[t]
\caption{Cross section per nucleon $\sigma_0$ or $BR\cdot\sigma_0$
for $\Xi^*$ production in $\Sigma^-$-nucleus interactions at
340{\gevc1}. The 3$\sigma$ upper limit for $\Xi^{--}(1860)$
production was determined in a mass bin of 20\Mevc2 width.
\label{tab:wa8902}}
\begin{ruledtabular}
\begin{tabular}{lccc}
 Particle, decay channel & $\sigma_0$ [$\mu$b]& $BR\cdot\sigma_0$ [$\mu$b]& Ref.\\
\colrule
$\Xi^-(1320)$ &1000$\pm$40~ & &\cite{WA89:xi}\\
$\overline{\Xi}^+(1320)$ &23$\pm$2& &\cite{WA89:v0}\\
$\Xi^0(1530)$ &218$\pm$44& &\cite{WA89:xistar}\\
$\Xi^0(1690) \rightarrow \Xi^-\pi^+$ & &2.5-6.8&\cite{WA89:ximpip,WA89:xistar}\\
$\Xi^0(1820) \rightarrow \Xi^0(1530)\pi^-$ & &21$\pm$5&\cite{WA89:xistar}\\
$\Xi^0(1950) \rightarrow \Xi^0(1530)\pi^-$ & &12$\pm$3&\cite{WA89:xistar}\\
$\Xi^{--}(1860) \rightarrow \Xi^-\pi^-$& & $\leq$ 3.1 (C) & this\\
 & & $\leq$ 3.5 (Cu)& work\\
\end{tabular}
\end{ruledtabular}
\end{table}

Limits on the integrated production cross sections $\sigma$ were
calculated by summing quadratically the contributions $ d\sigma
/dx_F \cdot \Delta x_F$ in the five individual $x_F$ bins listed
in column 4 of Tab. \ref{tab:wa8901}. The results are $BR\cdot
\sigma_{max}(0.15<x_F<0.9)$= 16 and 55 $\mu b$ per nucleus in case
of the carbon and copper target, respectively. An extrapolation to
the cross sections per nucleon yields the two values $BR\cdot
\sigma_{0,max} = 3.1\, \mu b$ for the carbon and $3.5\, \mu b$ for
the copper target, in excellent agreement with each other. As can
be seen from Tab. \ref{tab:wa8902}, these limits do not exceed the
production cross sections of all other observed $\Xi^*$
resonances.

At large $x_F$ a significant fraction of the $\Xi^-$ are produced
by interactions induced by the $\Xi^-$ beam contamination
\cite{WA89:xi,WA89:xipol}. Even if we were to assume that the
$\Xi^{--}(1860)$ production can be attributed exclusively to the
1.3\% $\Xi^-$ admixture in the beam, we obtain e.g. for the carbon
target and $x_F\geq$0.5 a limit for the $\Xi^{--}$ production by
$\Xi^-$ of 740$\mu$b. For comparison, even this large 3$\sigma$
limit corresponds to only 4\% of the $\Xi^-$ production cross
section in $\Xi^-$+Be interactions at 116{\gevc1} in the same
kinematic range \cite{Bia87}.

Finally we note that the $\Xi^- \pi^+$ mass distribution observed
by WA89 has already been published previously \cite{WA89:ximpip}
(see also Tab. \ref{tab:wa8902}). This combination is dominated by
the peak from $\Xi^0(1530)$ decays. The observed central mass was
in good agreement with the known value of M = 1531.8 $\pm$
0.3{\Mevc2} \cite{pdg}. Unfolding the observed width with the
width of the $\Xi_{1530}$ of $\Gamma = 9.1${\Mevc2} \cite{pdg}
gave an experimental resolution of $\sigma _{\Xi^{0}(1530)}$ =
3.7{\Mevc2}. Furthermore, a weak resonance signal with a width of
$\Gamma = 10 \pm 6 \Mevc2\ $ is visible at $M = 1686 \pm 4 \Mevc2\
$ above a large background. In the mass region of the
$\Xi^0(1860)$ (last three channels in the left part of Fig.~1a in
Ref. \cite{WA89:ximpip}) no enhancement over the uncorrelated
background can be seen in the WA89 data.

If the $\Xi^{--}$ signal observed by the NA49 collaboration is
real, then the non-observation in our experiment is not easily
understood. Generally particle ratios do not vary significantly
for the beam momentum range in question (160\gevc1 vs. 340\gevc1)
\cite{Let03,Liu04}. The fact that the $\Theta^+(1530)$ has been
seen in reactions on complex nuclei
\cite{Theta:NEUTRINO,Theta:SVD} makes also the different targets
(hydrogen vs. C, Si, Cu) an unlikely cause for the discrepancy.
The internal structure of the $\Sigma^-$ projectile or of the
$\Xi^{--}(1860)$ could be a more plausible reason for the rather
low limit of the $\Xi^{--}(1860)$/$\Xi^-$ ratio. It is well known,
that a transfer of a strange quark from the beam projectile to the
produced hadron enhances the production cross sections in
particular at large $x_F$ (see, for instance, Fig. 8 in
\cite{WA89:xi}). The different leading effects for octet and
decuplet  $\Sigma$ states \cite{WA89:sigma} even hint at an [sd]
diquark transfer from the $\Sigma^-$ projectile \cite{WA89:poc01}.
The production of a pentaquark containing correlated quark-quark
pairs (see e.g. Ref.\cite{Jaf03}) would probably benefit from such
a diquark transfer. However, for example in case of an extended
$\overline{K}-N-\overline{K}$ molecular structure of the
$\Xi(1860)$ \cite{Bic04} an [sd] diquark transfer may not
necessarily enhance the $\Xi^{--}$ production leading also to a
narrower $x_F$ distribution. As a consequence the cross section in
$\Sigma^-$ induced reactions might not exceed the one for
production in pp interactions. The latter cross section is
predicted to be $\sim$ 4$\mu$b \cite{Liu04} which is then close to
our limit.

Thus, if future high statistics experiments will confirm the
production of the $\Xi^{--}(1860)$ resonance in proton-proton
interaction, the non-observation with the $\Sigma^-$ beam may
point to a very exceptional production mechanism possibly related
to an exotic structure of the $\Xi^{--}(1860)$.



\begin{thebibliography}{19}
\bibitem{WA89:events}
Because of difficulties in reprocessing some of the old data
files, the $\Xi^-$ sample used in the present analysis is
approximately 25$\%$ smaller than the one used in a $\Xi^-$
polarization analysis of Ref. \cite{WA89:xipol}.
\end{thebibliography}
\end{document}